\documentclass[sigconf]{acmart} 
\usepackage{hyperref}

\AtBeginDocument{%
  \providecommand\BibTeX{{%
    \normalfont B\kern-0.5em{\scshape i\kern-0.25em b}\kern-0.8em\TeX}}}

\usepackage[font=small,skip=0pt]{caption}
\usepackage{balance}
\usepackage{booktabs}
\usepackage{tabularx}
\usepackage{enumitem}
\usepackage{ragged2e}
\usepackage{listings}
\usepackage{caption}

\usepackage{listings}
\usepackage{xcolor}
\usepackage{url}
\usepackage{amsmath}
\usepackage{xurl}  
\usepackage{hyperref}

\definecolor{codegray}{rgb}{0.55,0.55,0.55}
\definecolor{codeblue}{rgb}{0.00,0.25,0.70}
\definecolor{codepurple}{rgb}{0.50,0.00,0.55}
\definecolor{codebg}{rgb}{0.98,0.98,0.98}

\lstdefinestyle{kcwe}{
  backgroundcolor=\color{codebg},
  basicstyle=\ttfamily\scriptsize,
  numbers=left,
  numberstyle=\tiny\color{codegray},
  numbersep=4pt,
  xleftmargin=1.2em,
  frame=single,
  framerule=0.4pt,
  rulecolor=\color{codegray},
  breaklines=true,
  columns=fullflexible,
  showstringspaces=false,
  commentstyle=\color{codegray}\itshape,
  keywordstyle=\color{codeblue}\bfseries,
  stringstyle=\color{codepurple},
  tabsize=2,
  escapeinside={(*@}{@*)}
}
\copyrightyear{2026}
\acmYear{2026}
\setcopyright{cc}
\setcctype{by-nc-nd}
\acmConference[L@S '26]{Proceedings of the Thirteenth ACM Conference on Learning @ Scale}{June 29-July 03, 2026}{Seoul, Republic of Korea}
\acmBooktitle{Proceedings of the Thirteenth ACM Conference on Learning @ Scale (L@S '26), June 29-July 03, 2026, Seoul, Republic of Korea}
\acmDOI{10.1145/3774398.3811577}
\acmISBN{979-8-4007-2293-6/2026/06}

\begin{document}

\title[Personalized Worked Example Generation Using Pattern-based Knowledge Components]{Personalized Worked Example Generation from Student Code Submissions Using Pattern-based Knowledge Components}

\author{Griffin Pitts}
\orcid{0009-0004-3111-6118}
\affiliation{%
   \institution{North Carolina State University}
  \city{Raleigh}
  \state{NC}
  \country{USA}}
\email{wgpitts@ncsu.edu}
  
\author{Muntasir Hoq}
\orcid{0000-0003-2591-0476}
\affiliation{%
 \institution{North Carolina State University}
  \city{Raleigh}
  \state{NC}
  \country{USA}}
  \email{mhoq@ncsu.edu}

\author{Peter Brusilovsky}
\orcid{0000-0002-1902-1464}
\affiliation{%
   \institution{University of Pittsburgh}
  \city{Pittsburgh}
  \state{PA}
  \country{USA}}
  \email{peterb@pitt.edu}

\author{Narges Norouzi}
\orcid{0000-0001-9861-7540}
\affiliation{%
   \institution{University of California, Berkeley}
  \city{Berkeley}
  \state{CA}
  \country{USA}}
  \email{norouzi@berkeley.edu}

\author{Arto Hellas}
\orcid{0000-0001-6502-209X}
\affiliation{%
  \institution{Aalto University}
  \city{Espoo}
  \country{Finland}}
\email{arto.hellas@aalto.fi}

\author{Juho Leinonen}
\orcid{0000-0001-6829-9449}
\affiliation{%
  \institution{Aalto University}
  \city{Espoo}
  \country{Finland}}
\email{juho.2.leinonen@aalto.fi}
  
\author{Bita Akram}
\orcid{0000-0001-5195-5841}
\affiliation{%
   \institution{North Carolina State University}
  \city{Raleigh}
  \state{NC}
  \country{USA}}
  \email{bakram@ncsu.edu}

\renewcommand{\shortauthors}{Griffin Pitts et al.}

\begin{abstract}
Adaptive programming practice often relies on fixed libraries of worked examples and practice problems, which require substantial authoring effort and may not correspond well to the logical errors and partial solutions students produce while writing code. As a result, students may receive learning content that does not directly address the concepts they are working to understand, while instructors must either invest additional effort in expanding content libraries or accept a coarse level of personalization. We present an approach for knowledge-component (KC) guided educational content generation using pattern-based KCs extracted from student code. Given a problem statement and student submissions, our pipeline extracts recurring structural KC patterns from students' code through AST-based analysis and uses them to condition a generative model. In this study, we apply this approach to worked example generation, and compare baseline and KC-conditioned outputs through expert evaluation. Results suggest that KC-conditioned generation improves topical focus and relevance to students’ underlying logical errors, providing evidence that KC-based steering of generative models can support personalized learning at scale.
\end{abstract}

\ccsdesc[500]{Social and professional topics~Computing education}
\ccsdesc[300]{Computing methodologies~Natural language processing}
\ccsdesc[300]{Computing methodologies~Knowledge representation and reasoning}
\ccsdesc[300]{Computing methodologies~Information extraction}

\keywords{Worked example generation; programming education; knowledge components; AST-based code analysis; LLMs; AI in education}

\maketitle

\section{Introduction}

Introductory programming courses often use short learning activities that support students' conceptual understanding through practice and explanation. One common activity type is the worked example, which typically presents students with a related problem statement, a code solution similar to what they are working on, and step-by-step instructional explanations that walk through the solution process \cite{skudder2014workedexamples,muldner2022review}. Access to a well-matched worked example at points of struggle can help students interpret errors, connect ideas to code, and keep moving forward \cite{hoq2025automated}. Grounded in cognitive load theory, worked examples reduce unnecessary load by making intermediate steps explicit and limiting the inference required to follow a solution \cite{sweller1988cognitive,chandler1991cognitive,muldner2022review}.

Producing high-quality worked examples at the level of detail needed for instruction can be costly and time-intensive. Instructors must craft correct solutions, decide how to segment them into meaningful steps, and write explanations appropriate for novice programmers \cite{smit2025personalising,jury2024evaluating}. Large repositories can reduce the need to author every example from scratch, but identifying examples that align with a student’s specific task and difficulty remains challenging \cite{hoq2025automated}. In many cases, the repository may not contain a worked example with the needed fine-grained relevance, and students’ prior knowledge further complicates selection because explanatory detail that helps novices can become redundant for advanced learners \cite{kalyuga2003expertisereversal}.

Large language models (LLMs) offer a way to reduce authoring burden by generating worked examples on demand \cite{jury2024evaluating,pitts2025surveyllm}. Prior evaluations in introductory programming report that expert reviewers judged LLM-generated worked examples to have clear, suitable explanations and coherent step structure, and that students found them helpful for learning and making progress on practice tasks \cite{jury2024evaluating,smit2025personalising}. However, existing approaches often provide limited adaptation to learner differences such as skill level, which can lead to examples that introduce unfamiliar concepts for novices \cite{smit2025personalising}. They may also fail to align with the specific logical error reflected in a student’s partial code solution \cite{hoq2025logical}. To make generated worked examples more directly tied to what students appear to be struggling with in code, we use knowledge components as targets for generation.

Knowledge components (KCs) represent discrete units of knowledge or skill that support successful task performance \cite{koedinger2012knowledge}. In this work, we operationalize KCs as the concepts that a worked example should conceptually target. Specifically, following prior work \cite{2508.09281v2}, we extract \textit{pattern-based KCs} from a student’s submission via identifying common programming patterns in abstract syntax trees (AST) found in a cohort of students' submissions. As these KCs are derived from recurring AST substructures in the student’s own code, they ground generation in constructs students are actively using and may be applying incorrectly or incompletely \cite{2508.09281v2}. We then condition an LLM to generate a worked example whose code and explanations explicitly address these KC targets.

We compare the KC-conditioned approach to an otherwise identical baseline that omits KC information. Both approaches are evaluated through expert review using a rubric adapted from prior work \cite{jury2024evaluating}. For outputs guided by the pattern-based KCs, we additionally assess whether the intended KCs appear in the example code and are explicitly discussed in the step explanations. Altogether, the following research questions are addressed: \begin{enumerate}[leftmargin=*,label=\textbf{RQ\arabic*:}]
\item Does KC-conditioned generation improve worked-example quality and relevance compared to our baseline?
\item How reliably does our KC-conditioned pipeline produce worked examples that include and explicitly explain the intended KC patterns?
\end{enumerate}

\section{Related Work}

Recent work has used LLMs to generate programming learning activities. Focusing on scaffolded Parsons problems, CodeTailor \cite{hou2024codetailor} uses a student's incorrect code to prompt an LLM for a corrected reference solution, then constructs a personalized Parsons problem by aligning the student's original code with the generated correction. Similarly, PuzzleMakerPy \cite{del2024automating} uses an LLM to generate Parsons problems that learners can customize by thematic context and targeted concepts. For LLM-generated worked examples, prior work has examined how prompting strategies affect content quality. Jury et al. \cite{jury2024evaluating} studied worked example generation for CS1 students with both expert and student evaluation, reporting that LLMs can produce generally coherent worked examples, while also noting LLM outputs often lacked novice-friendly detail in step structure and commenting. Related work has evaluated similar generation in a CS2 context \cite{lindvall2024using} and explored controllable rewriting that produces novice, intermediate, and advanced variants from a base worked example by adjusting step granularity and explanation detail \cite{smit2025personalising}. In this work, we extend this line of research by conditioning LLM-based worked-example generation on pattern-based KCs extracted from student code submissions, linking generation to an explicit representation of a learner's likely logical error.

\section{Methodology}

Our methodology has three stages. First, we extract pattern-based KCs from student submissions and enrich them with short, human-readable descriptions. Second, for each submission we generate two worked examples using a baseline prompting pipeline and a KC-conditioned prompting pipeline. Third, we evaluate the generated worked examples through expert review.

\subsection{Dataset and Sampling}
We use a publicly available CodeWorkout dataset containing 57{,}670 anonymized Java submissions from 368 CS1 students collected in Spring 2019, including 18{,}787 correct and 38{,}883 incorrect submissions \cite{edwards2017codeworkout}. For this study, we randomly selected two out of fifty problems from the dataset, \texttt{repeatEnd} and \texttt{fix45} \footnote{\url{https://codeworkout.cs.vt.edu/gym/exercises}}. \texttt{repeatEnd} is a string task involving substring extraction and repetition (repeat the last $n$ characters $n$ times), while \texttt{fix45} is an array task involving constrained rearrangement (reorder elements so each 4 is immediately followed by a 5 without moving the 4s). For each problem, we identify a student’s last incorrect attempt by timestamp, since incorrect submissions provide the clearest evidence of students' code patterns involving logical errors our worked-example generation pipeline was designed to address. From these candidate submissions, we randomly sample 50 submissions per problem.

\subsection{Pattern-based KC Discovery and Labeling}
We adopt the pattern-based KC discovery pipeline of Hoq et al. \cite{2508.09281v2} to derive submission-level KC targets from student code. The pipeline operationalizes a KC as a recurring AST-subtree pattern that corresponds to a programming construct or a common combination of constructs that appears across many student solutions. We run the pattern-based KC discovery pipeline on the full Codeworkout dataset following the procedure in \cite{2508.09281v2}, and use the resulting trained components to infer KC targets for the sampled submissions.

\paragraph{Step 1: Identify important subtrees with attention.}
For each submission, we parse the program into an AST and extract candidate subtrees ranging from small local constructs to larger statement-level structures. These subtrees are fed to a Subtree-based Attention Neural Network (SANN) trained to predict whether a submission is correct or incorrect. SANN encodes each subtree into a fixed-length vector and assigns an attention weight indicating its importance for the prediction \cite{2508.09281v2}. Following Hoq et al. \cite{2508.09281v2}, attention weights are computed with a sigmoid activation so multiple subtrees can be highlighted within a program, identifying influential patterns in the code responsible for correctness. 

\paragraph{Step 2: Normalize subtree tokens for abstraction.}
We normalize each retained subtree by replacing identifiers and literals with placeholders \cite{2508.09281v2}. This reduces sensitivity to surface-level variation and supports grouping similar patterns across student implementations.

\paragraph{Step 3: Learn representations for clustering.}
We train a variational autoencoder (VAE) over sequences of high-attention subtrees to learn representations that reflect how each subtree appears in program context \cite{2508.09281v2}. The VAE is trained on high-attention subtrees from correct submissions so that structurally similar correct patterns embed nearby in the latent space \cite{2508.09281v2}. After training, we use the encoder’s latent vectors as inputs to clustering.

\paragraph{Step 4: Cluster latent vectors into a KC inventory, then map submissions to KCs.}
We cluster the VAE latent vectors from correct submissions using K-means (k = 50) \cite{2508.09281v2}. At inference time, each high-attention subtree from either a correct or incorrect submission is assigned to its nearest cluster centroid to obtain a KC ID.

\paragraph{Step 5: Enrich KC targets for prompting.}
Because cluster IDs, generated following~\cite{2508.09281v2}, are not human-readable, we add an enrichment layer that attaches short labels and one-sentence descriptions to each submission’s KC targets for use in prompts. For each KC assigned to a submission, we provide GPT-5.2 (\textit{gpt-5.2-chat-latest}) with (i) the problem statement, (ii) the student submission, and (iii) a code snippet aligned to the high-attention subtree that triggered that KC assignment, then prompt it to produce a concise KC label (2--6 words) and a one-sentence description of the pattern.

\subsection{Worked Example Generation}

With each student submission, we derive a submission-level KC target set from the code using the Hoq et al. \cite{2508.09281v2} pipeline described in Section 3.2, and generate human-readable labels and a single-sentence description for every pattern-based KC. Using these KCs, we generate two worked examples per student submission with GPT-5.2 (\textit{gpt-5.2-chat-latest}), one generated with the KC targets included in the prompt and one generated that omits KC information. 

\paragraph{Prompting and problem context.} The two generation pipelines (baseline and KC-conditioned) use a shared prompt template that specifies a persona as an introductory programming tutor and lays out the task, response structure, and formatting guardrails. In both variants, the prompt provides the problem statement and the student’s code, and instructs the model to infer what the student is struggling with and generate a worked example for a related but different problem that practices similar skills. The KC-conditioned variant additionally provides the student submission’s extracted pattern-based KCs, as labels with single-sentence descriptions, and instructs the model to use these KCs as constraints when inferring the student’s logical error and selecting what the worked example should practice and explain. Both pipelines enforce the same output format, based on prior work \cite{jury2024evaluating}, where the LLM is instructed to produce a worked example with 3--10 steps, and each step pairs a brief explanation with a corresponding code fragment \footnote{Prompt templates: \url{https://osf.io/4h9dn/overview?view_only=97189a73e56b4254bd2298669b6eabc4}}.

\subsection{Expert Evaluation}
Experts evaluated baseline and KC-conditioned outputs using the rubric outlined in Table~\ref{tab:rubric}, adapted from prior work \cite{jury2024evaluating}. Each rubric item is scored on a 0--2 scale, where 0 indicates the item criteria is not met, 1 indicates partial fulfillment (meets the item criteria in some aspects but with omissions), and 2 indicates the item criteria is fully met. For KC-conditioned outputs, experts additionally score KC coverage, which reflects whether the provided KC targets are instantiated in the worked example code and step explanations. Further, for each paired comparison, experts recorded their preference for the baseline or KC-conditioned worked example, or indicated no preference. Per student submission sampled, experts are shown the problem statement, the student’s code, and the paired worked examples. For worked examples generated with pattern-based KCs, the KC labels and descriptions are also shown. 

The evaluation followed a procedure recommended by \cite{landis1977measurement}. Two experts first jointly coded an initial subset of generated worked examples (10\%), then computed Cohen’s Kappa ($\kappa$) on that subset to quantify agreement. Inter-rater agreement exceeding $\kappa$ = 0.80 was achieved on the first round ($\kappa$ = 0.90), after which the full set of generated worked examples was coded and analysis conducted. Paired Wilcoxon signed-rank tests and Holm corrected p-values were used when comparing baseline and KC-conditioned ratings.

\begin{table}[t]
\centering
\small
\setlength{\tabcolsep}{6pt}
\renewcommand{\arraystretch}{1.2}
\begin{tabularx}{\linewidth}{@{}p{0.25\linewidth}X@{}}
\toprule
\textbf{Item} & \textbf{Description} \\
\midrule
Formatting &
Does the output follow the required worked-example structure (3--10 steps), where each step pairs code with a matching written explanation and includes helpful inline \texttt{//} comments? \\
Clear explanations &
Is the worked example easy to follow, with a clear, linear flow and little assumed prior knowledge? \\
Correctness &
Is the code correct by inspection, and would it produce the intended behavior for the problem? \\
Step structure &
Are the steps broken into manageable parts, with each step covering a specific part of the solution and avoiding large jumps? \\
Relevance to the student &
Does the worked example address a misconception or logical error suggested by the student submission? \\
KC coverage &
Are the provided KC targets present in the code and addressed in the step text in a way that clearly connects each target to where it appears? \\
\bottomrule
\end{tabularx}
\caption{Expert-evaluation rubric for worked examples.}
\label{tab:rubric}
\end{table}

\section{Results}

\paragraph{RQ1: Worked example quality and relevance.}
Across 100 generated worked examples (50 baseline, 50 KC-conditioned), experts scored outputs on five rubric items, scaled 0--2 (Table~\ref{tab:rubric}). For KC-conditioned outputs, experts also rated a sixth item, KC coverage. The largest differences appeared on two items: novice-friendly explanations and student relevance (Table 2). KC-conditioned worked examples scored higher on student relevance (22\% higher; $p=0.001$), indicating that they more often addressed the logical error suggested by the student submission rather than offering a generic solution. KC-conditioned outputs also scored higher on novice-friendly explanations (13\% higher; $p=0.027$), which aligns with our observations that their step text more consistently explains the KC-targeted parts of the solution at a level of detail appropriate for novices. The remaining rubric items showed no statistically significant differences. Both pipelines consistently followed the intended worked example format and produced correct code at near perfect levels. However, one tradeoff we observed was that KC-conditioned outputs received slightly lower ratings in regard to their step-structure (5\% lower; $p=0.076$), which may reflect cases where incorporating additional KC targets leads the model to pack more content into each step, making step boundaries less distinct and step labels less informative. The preferences of the expert evaluators were considered, and were observed to be consistent with the rubric item ratings. Across 100 comparisons, experts had no preference for 50, preferred the KC-conditioned worked examples for 41, and preferred the baseline for 9. Based on our observations, preferences for the KC-conditioned output were typically tied to the outputs being more specific toward what the student likely misunderstood and more direct in addressing that gap, instead of presenting a broadly correct but generic solution.

\begin{table}[t]
\centering
\footnotesize
\setlength{\tabcolsep}{4pt}
\renewcommand{\arraystretch}{1.1}
\begin{tabular}{lrrr}
\toprule
\textbf{Rubric item} & \textbf{Baseline $M$} & \textbf{KC-conditioned $M$} & \textbf{$p$-value} \\
\midrule
Formatting        & 2.00 & 2.00 & 1.00 \\
Clear explanations & 1.81 & 1.94 & 0.027 \\
Correctness                 & 1.97 & 1.99 & 0.635 \\
Step Structure             & 1.99 & 1.94 & 0.076 \\
Relevance to the student             & 1.67 & 1.89 & 0.001 \\
\bottomrule
\end{tabular}
\caption{Baseline vs.\ KC-conditioned ratings (0--2 scale). $M$ denotes the mean expert rating across 100 generated worked examples.}
\label{tab:paired_results}
\end{table}

\paragraph{RQ2: KC Coverage.} For KC-conditioned outputs, we asked whether the provided KC targets appear in the code and are explained in the step text in a way that link each target to the code where it appears. Alignment was high (mean $=1.92$ on a 0--2 scale), and of 100 examples, 92 fully met this criterion, with 8 partially aligned. Partial alignment occurred primarily when multiple KC targets were provided, where not all targets were addressed clearly in the code and step explanations. In these cases, the worked example often focused on one KC and mentioned others briefly, and step boundaries were sometimes less distinct when the model addressed several targets at once. This suggests prompts may need additional guidance for prioritizing multiple KC targets, including which targets to emphasize and how to allocate them across steps.

\begin{figure}[t]
\centering

\textbf{Student code excerpt (problem \texttt{fix45}, code\_id=43552)}
\begin{lstlisting}[style=kcwe,language=Java]
if (i == 0 && nums[i] == 5 || nums[i] == 5 && nums[i-1] != 4) {
    int fiveSpot = i;
    for (int m = i; m < nums.length; m++) {
        if (nums[m] == 4 && nums[m+1] != 5) {
            int otherNum = nums[m+1];
            nums[m+1] = 5;
            nums[fiveSpot] = otherNum;
            break;
        }}}
\end{lstlisting}

\textbf{Extracted pattern-based KC (label + description)}
\begin{lstlisting}[style=kcwe]
Pattern-based KC from the student submission: (*@\textbf{Boolean operator precedence}@*)

Description: The student combines (*@\textbf{\texttt{\&\&}}@*) and (*@\textbf{\texttt{||}}@*) in a single condition without parentheses, so the intended grouping may differ from Java's evaluation order.
\end{lstlisting}

\textbf{KC-conditioned worked example excerpt}
\begin{lstlisting}[style=kcwe,language=Java]
Question: Write a method that counts how many numbers in an array are either negative OR are zeros that do not come immediately after another zero. Be careful to correctly combine && and || in your condition.

Overview: This program will loop through an integer array and count values that meet a combined condition using both && and ||. We will use parentheses to make the logic clear and avoid mistakes with operator precedence.

Step 1: Create the method and a counter variable.
public int countSpecial(int[] nums) {
    int count = 0;                  // Keeps track of how many values meet the condition

Step 2: Loop through the array so we can check each element.
    for (int i = 0; i < nums.length; i++) {   // Go through every index

Step 3: Use parentheses to clearly group && and || conditions.
        // We want to count: 1) Any negative number OR
        // 2) A zero that is NOT immediately after another zero
        if (nums[i] < 0 || (nums[i] == 0 && (i == 0 || nums[i - 1] != 0))) {
            count++;                // Increase count if condition is true
        }

Step 4: Finish the loop and return the result.
    }
    return count;                   // Return the final total
}
\end{lstlisting}

\caption{Example of a high-level pattern-based KC extracted from student code and used to guide worked-example generation, targeting explicit grouping of operations with parentheses.}
\label{fig:kc_guided_example_43552}
\end{figure}

\section{Discussion and Future Work}

This work proposes a KC-guided worked example generation pipeline that uses pattern-based KCs extracted from student code (via AST subtree discovery) as constraints in the generation process. By conditioning generation on these KCs, the resulting worked examples are designed to practice and explain patterns associated with students’ likely logical errors. Across expert ratings, the KC-conditioned pipeline significantly improved evaluations of explanation clarity and relevance to the student’s likely logical error. This result suggests that the use of KCs as a structured representation of what a learner needs to practice can support more targeted personalization of generated learning content. However, the partial KC-coverage cases also suggest that KC-guided generation needs mechanisms for deciding which targets to emphasize when a submission maps to several plausible KCs. In future work, these KC targets could be integrated into a broader learner modeling framework that tracks students’ mastery over time and uses that knowledge state to select and sequence KC-guided examples across practice opportunities.

Future work will prioritize establishing potential learning impacts of our KC-conditioned worked examples. While expert evaluations indicate that KC inclusion can steer LLM-generation toward clearer and more instructionally-relevant instruction, an important question remains whether these examples improve learning and transfer. We plan future classroom studies where an incorrect submission triggers either a baseline or a KC-guided worked example, and we will measure improvement on a second attempt, performance on a related follow-up problem targeting the same KCs, and brief concept checks aligned with those targets. A limitation of our approach is that KC-guided generation quality depends on the quality and granularity of the discovered KC inventory and its labels, and on how accurately a new submission is mapped to the most instructionally-relevant targets. This can be challenging when a submission yields many plausible KC targets. We plan to address this by incorporating model attention weights in the generation process to prioritize KC targets, validating the human-readable KC descriptions with instructors, and testing robustness across additional problems and programming topics.

\begin{acks} 
This research was supported by the U.S. National Science Foundation (NSF) under Grant \#2426837. Any opinions, findings, and conclusions expressed in this material are those of the authors and do not necessarily reflect views of the NSF.
\end{acks} 

\balance
\bibliographystyle{ACM-Reference-Format}
\bibliography{base}

\end{document}